\documentclass[useAMS,usenatbib]{mn2e}
\usepackage{times}
\usepackage{graphics,epsfig}
\usepackage{amssymb}

\newcommand{\kms}{km~s$^{-1}$}

\newcommand{\cm}{cm$^{-2}$}
\newcommand{\NHI}{N_{\rm HI}}

\newcommand{\PKSA}{PKS~1814$-$637}
\newcommand{\PKSB}{PKS~0407$-$658}

\title[The temperature of the WNM in the Milky Way]
{The temperature of the WNM in the Milky Way}
\author[Kanekar et al.]{Nissim Kanekar$^{1}$
\thanks{E-mail: nissim@astro.rug.nl (NK); rsubrahm@atnf.csiro.au (RS); 
chengalu@ncra.tifr.res.in (JNC); vickis@physics.usyd.edu.au (VS)}
Ravi Subrahmanyan$^{2}$, Jayaram N Chengalur$^{3}$, 
Vicky Safouris$^{4}$ \\
$^{1}$ Kapteyn Institute, University of Groningen, 
Post Bag 800, 9700 AV Groningen\\
$^{2}$ Australia Telescope National Facility, CSIRO,
Locked bag 194, Narrabri, NSW 2390, Australia  \\
$^{3}$ National Centre for Radio Astrophysics, 
Post Bag 3, Ganeshkhind, Pune 411 007 \\
$^{4}$ Dept. of Physics, University of Sydney, NSW 2006, Australia }
\begin{document}
\date{Received mmddyy/ accepted mmddyy}

\maketitle

\label{firstpage}

\begin{abstract}

We report high spectral resolution Australia Telescope 
Compact Array HI 21~cm observations resulting in the detection of 
the warm neutral medium (WNM) of the Galaxy in absorption against 
two extragalactic radio sources, \PKSA~ and \PKSB. 
The two lines of sight were selected on the basis of the simplicity 
of their absorption profiles and the strength of the background 
sources; the high velocity resolution of the spectra then enabled 
us to estimate the kinetic temperatures of the absorbing gas by 
fitting multiple Gaussians to the absorption profiles. Four separate 
WNM components were detected toward the two sources, with peak 
optical depths $\tau_{\rm max} = (1.0 \pm 0.08) \times 10^{-2}$, 
$(1.4 \pm 0.2) \times 10^{-3}$, $(2.2 \pm 0.5) \times 10^{-3}$ 
and $(3.4 \pm 0.5) \times 10^{-3}$ and kinetic temperatures 
$T_{\rm k} = 3127 \pm 300$~K, $3694 \pm 1595$~K, $3500 \pm 1354$~K 
and $2165 \pm 608$~K respectively. All four components were thus 
found to have temperatures in the thermally unstable range 
$500 < T_{\rm k} < 5000$~K; this suggests that thermal equilibrium 
has not been reached throughout the WNM.
\end{abstract}


\section{Introduction}
\label{sec:intro}

A corner stone of models for the Galactic interstellar medium (ISM) is 
that neutral hydrogen (HI) exists in two stable phases, in pressure 
equilibrium with one another \citep{field69} and with the hot ionized 
medium \citep{mckee77}. Observationally, it has been established that 
HI indeed has two phases, (1)~a cold dense phase (the cold neutral 
medium, CNM), which has high 21~cm optical depth and gives rise to the 
narrow absorption features seen toward continuum sources, and (2)~a 
warm diffuse phase, the warm neutral medium (WNM) which contributes to 
the emission, but is extremely difficult to detect in absorption due to 
its low optical depth.  Decades of study have established that the 
CNM has temperatures in the range $\sim 40 - 200$~K and number densities 
$n \sim 1 - 10$~cm$^{-3}$~ (e.g. \citealt*{dickey78,payne83,heiles03a}). 
On the other hand, while theoretical models (e.g. \citealt{wolfire95}) 
suggest that kinetic temperatures in the WNM lie in the range 
$\sim 5000-8000$~K, with number densities $\sim 0.01 - 0.1$~cm$^{-3}$, 
very little is as yet observationally known about physical conditions 
in this important constituent of the interstellar  medium 
(e.g. \citealt{kulkarni88}).

Temperature measurements in HI clouds are usually carried out by comparing 
the 21~cm optical depth in a given direction (obtained through 21~cm 
absorption studies toward background continuum sources) with the emission 
brightness temperature from nearby directions. This yields the excitation 
temperature of the HI gas, usually referred to as the ``spin temperature'', 
$T_{\rm s}$. In the CNM, $T_{\rm s}$ is driven toward the kinetic 
temperature $T_{\rm k}$ of the cloud, both by collisions and resonant 
scattering of Ly-$\alpha$ photons \citep{field58}. Thus, for the CNM, 
21~cm absorption/emission studies directly yield the kinetic temperature 
of an HI cloud or, in the case of multiple, blended, optically thin clouds 
along the line of sight, the column density weighted harmonic mean of the kinetic 
temperatures of the different clouds (e.g. \citealt{kulkarni88}). On the other hand,
the temperature of the WNM is still weakly constrained due to the difficulties
in detecting it in absorption (e.g. \citealt{mebold75,kulkarni85}). Further, 
the particle and Ly-$\alpha$ number densities in the WNM may be too low to 
thermalize the hyperfine levels; 
$T_{\rm s}$ could hence be significantly lower than 
$T_{\rm k}$~here \citep{field58,liszt01}.  21~cm absorption/emission 
studies of WNM clouds thus provide a {\it lower} limit to the 
kinetic temperature, even in the rare cases of claimed detections of the WNM 
(e.g. \citealt*{carilli98}).

A serious problem with $T_{\rm s}$ measurements via the classical 21~cm 
absorption/emission studies is that such studies involve the comparison of 
the HI optical depth along a given line of sight with the brightness 
temperature obtained from other directions, i.e. the assumption that the 
HI cloud is uniform on scales much larger than the beam size. This is 
often untrue for the CNM and may well be incorrect for the WNM. Additionally, 
in cases where such studies have been carried out with 
single-dish radio telescopes, the on-source absorption spectrum contains 
a contribution from HI emission in the beam; it is difficult to accurately 
correct for this effect. Single dish studies also require assumptions about 
the spatial distribution of HI clouds along the line of sight (which is 
unknown {\it a~priori}), to correct for absorption of background HI emission 
by foreground CNM. Further, emission measurements suffer from the problem 
of stray radiation entering via the side-lobes of the telescope beam.
Finally, searches for the WNM in absorption are usually confused by the 
multitude of CNM lines in any given direction, as even low column density 
CNM often has a higher optical depth than warm HI with a much higher column 
density.

We emphasize that it is the HI {\it emission} spectra which are most seriously 
affected by the above issues, stemming from stray radiation, non-uniformity 
of HI clouds across the beam, self-absorption, etc. These make it very
difficult to estimate the spin temperature of the WNM in the standard 
absorption/emission searches. Conversely, HI absorption studies toward
compact sources trace narrow lines of sight through the intervening clouds; 
when carried out using long baseline interferometers, these studies can 
resolve out the foreground HI emission (thus avoiding the above difficulties) 
and provide an uncontaminated measure of the absorption profile, which might
then be inspected for WNM features. The only problems with such attempts to 
detect the WNM are that (1)~the WNM optical depth is very low, i.e.  high 
sensitivity is necessary to detect it in absorption, and (2)~the WNM must 
be searched for in the midst of strong CNM absorption features. The latter 
issue can be mitigated by choosing lines of sight with simple CNM structure 
(with only a few narrow absorption components) and using high velocity 
resolution observations to model the deep narrow CNM features (e.g. as 
Gaussians) and subtract them out. One can then search for wide, shallow 
WNM absorption in the residuals and thus estimate (or constrain) the WNM 
temperature. Further, it is not necessary to know the spatial distributions 
of the absorbing clouds since optical depths are additive (for small $\tau$) 
and the HI emission is resolved out. In fact, the WNM has indeed been detected 
by the above approach (albeit at cosmological distances), in the $z = 0.0912$ 
and $z = 0.2212$ damped Lyman-$\alpha$ systems toward QSO B0738+313. 
\citep{lane00,kanekar01}. 

When the present observations were being planned, the Australia Telescope 
Compact Array (ATCA) was the only radio interferometer which could provide 
the requisite high velocity resolution ($\sim 0.5$~\kms), along with 
large bandwidths to search for broad absorption. We hence carried out a pilot
ATCA 21~cm absorption/emission survey toward a number of strong compact sources, 
to select those with the simplest absorption and emission profiles. Of these, 
PKS~0407$-$658 and PKS~1814$-$637 \citep{rad72} were chosen as the best candidates 
for a search for the WNM, both due to their high flux densities ($S_{20cm} \sim 15$~Jy) 
and simple profiles. 
The final ATCA observations and data analysis are described in Section \ref{sec:obs} and
the absorption spectra and Gaussian fits presented in Section \ref{sec:spectra};
finally, implications for the temperatures in the absorbing HI clouds are
discussed in Sections~\ref{sec:temp} and \ref{sec:discuss}.

\vskip -0.1in
\section{Observations and Data Analysis}
\label{sec:obs}

\setcounter{figure}{0}
\begin{figure*}
\hskip -5.2in
\epsfig{file=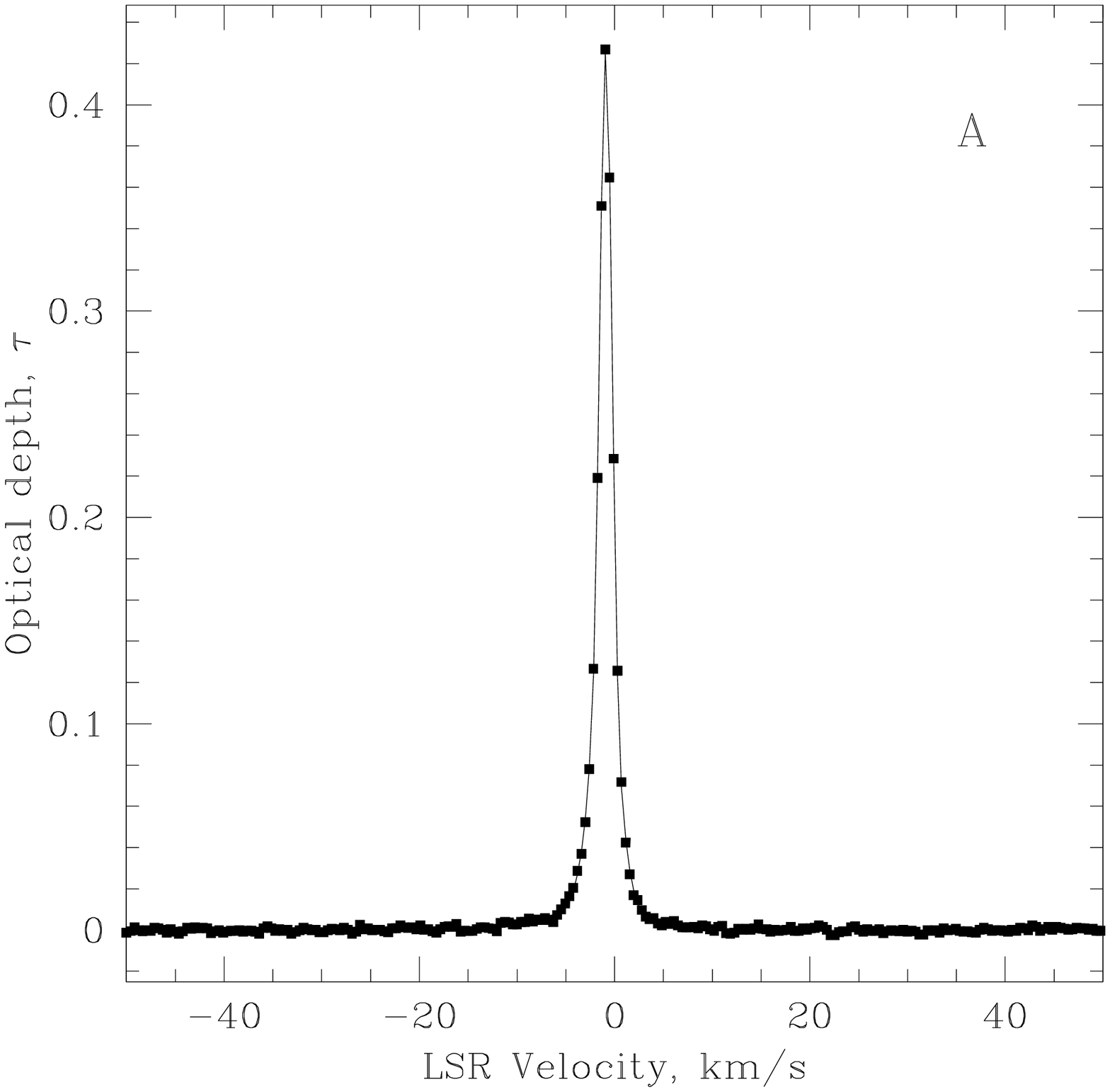,width=2.4in, height=2.4in}
\vskip -2.4 in
\hskip -0.3in
\epsfig{file=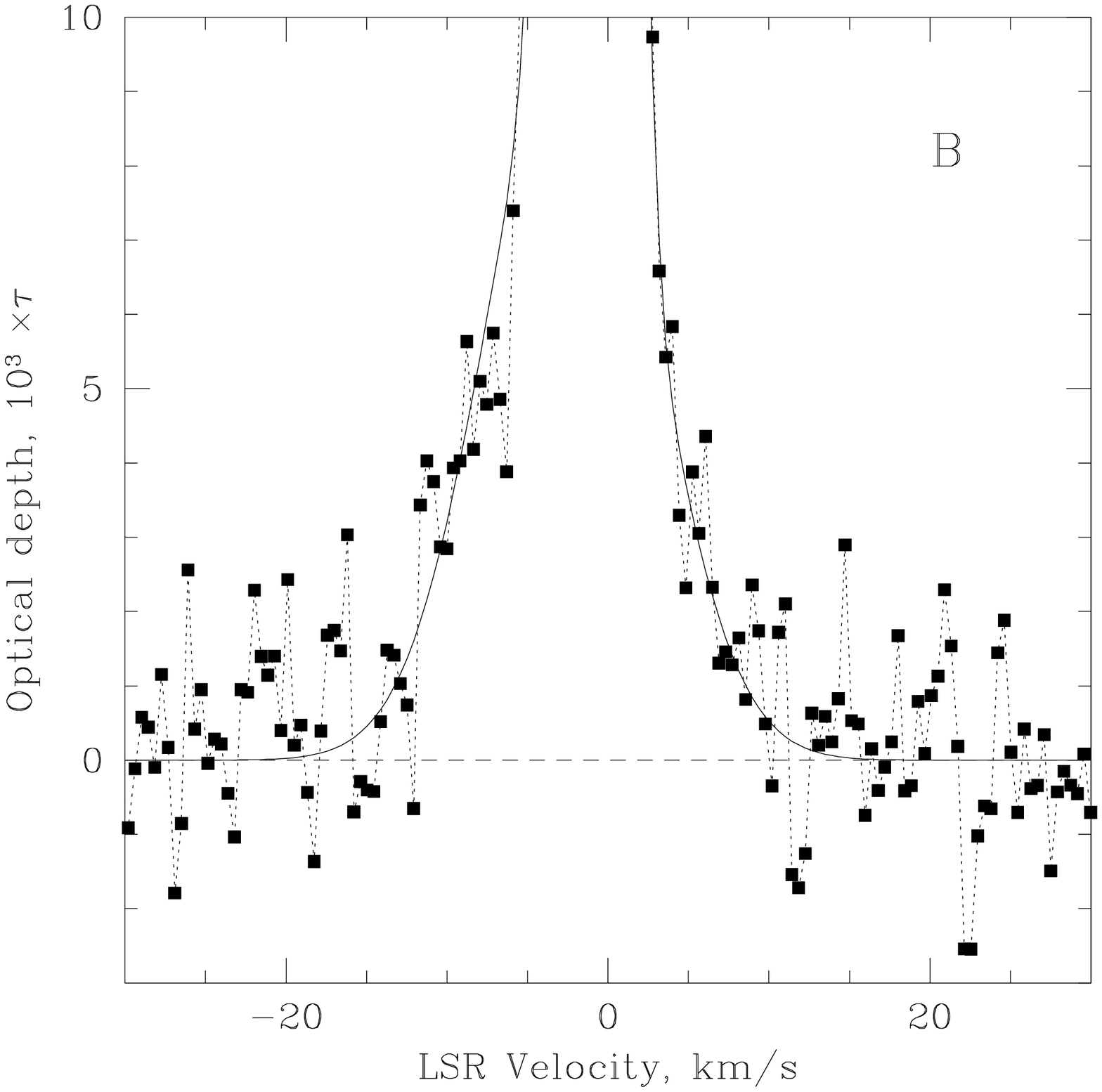,width=2.4in,height=2.4in}
\vskip -2.4 in
\hskip +4.5 in \epsfig{file=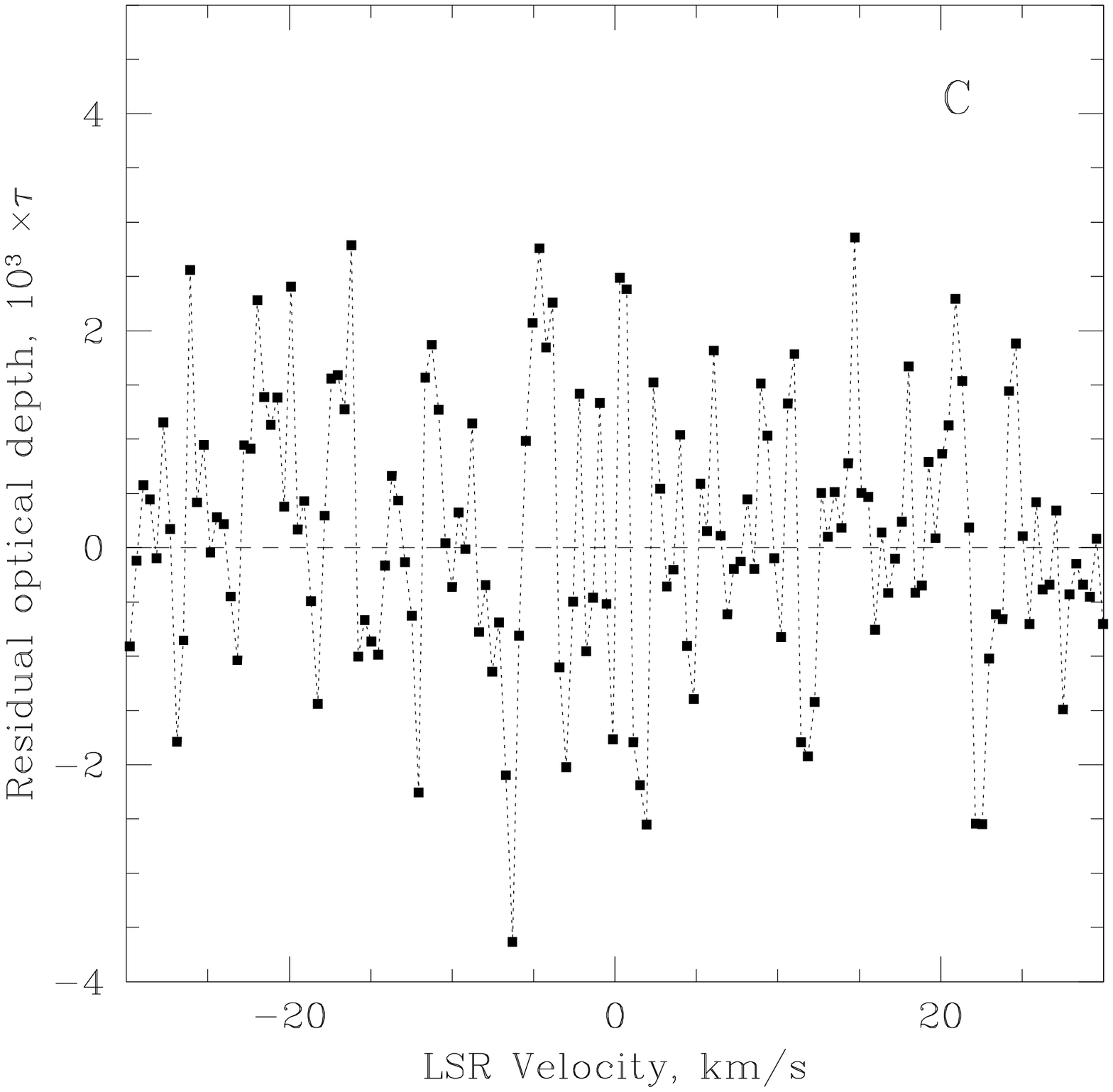,width=2.4in,height=2.4in}
\caption{ [A] : 0.4~\kms~resolution absorption spectrum toward \PKSA~(solid 
	      	points), along with the 3-Gaussian fit (solid line).
          [B] : A zoomed-in version of this spectrum. A wide absorption 
	  	component can be seen at the base of the narrow CNM features. 
          [C] : The residual absorption spectrum after subtracting the fit;
		the residuals are seen to lie within the noise.}
\label{fig:1814}
\end{figure*}

\setcounter{figure}{1}
\begin{figure*}
\hskip -5.2in
\epsfig{file=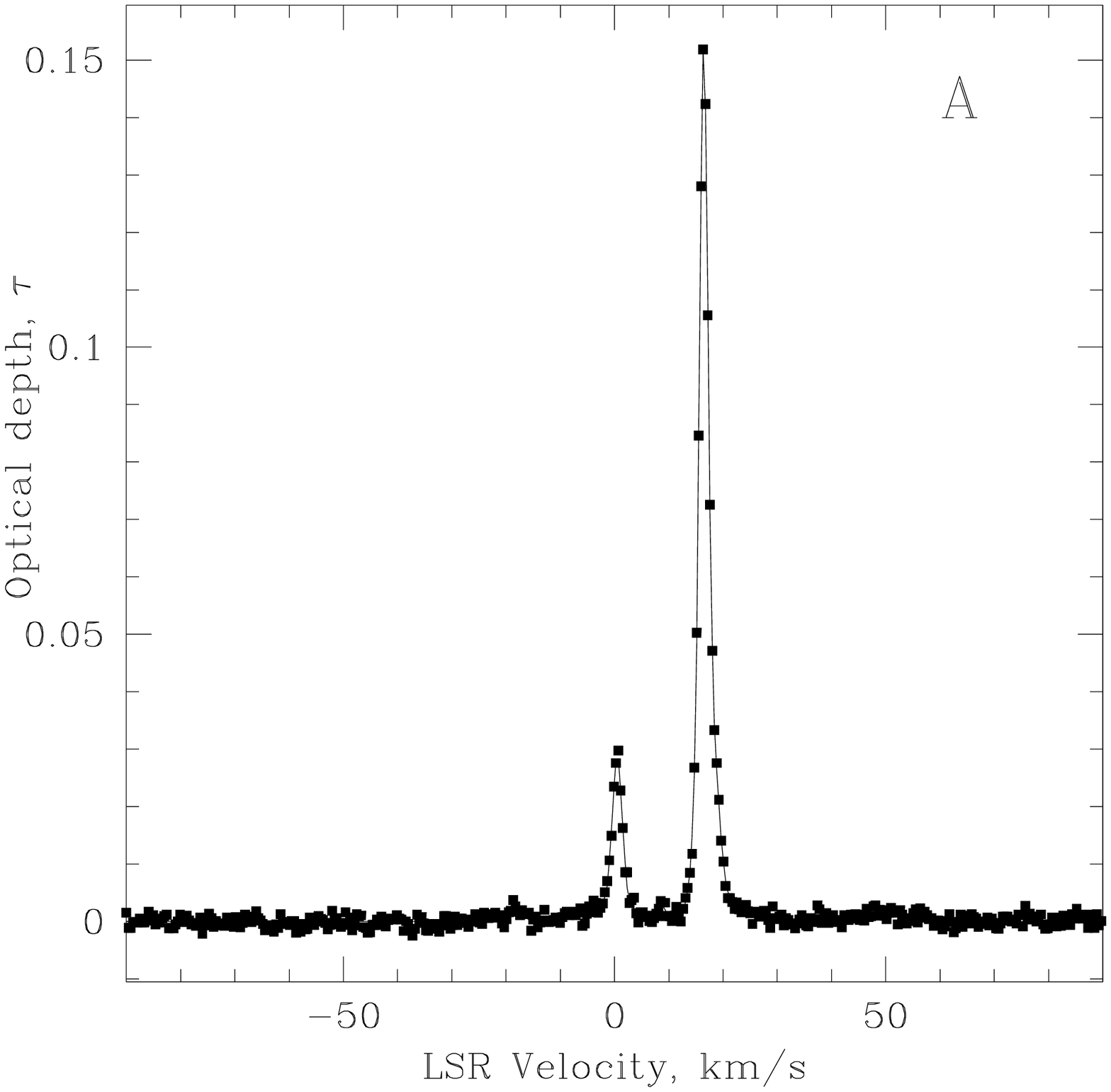,width=2.4in, height=2.4in}
\vskip -2.4 in
\hskip -0.3 in \epsfig{file=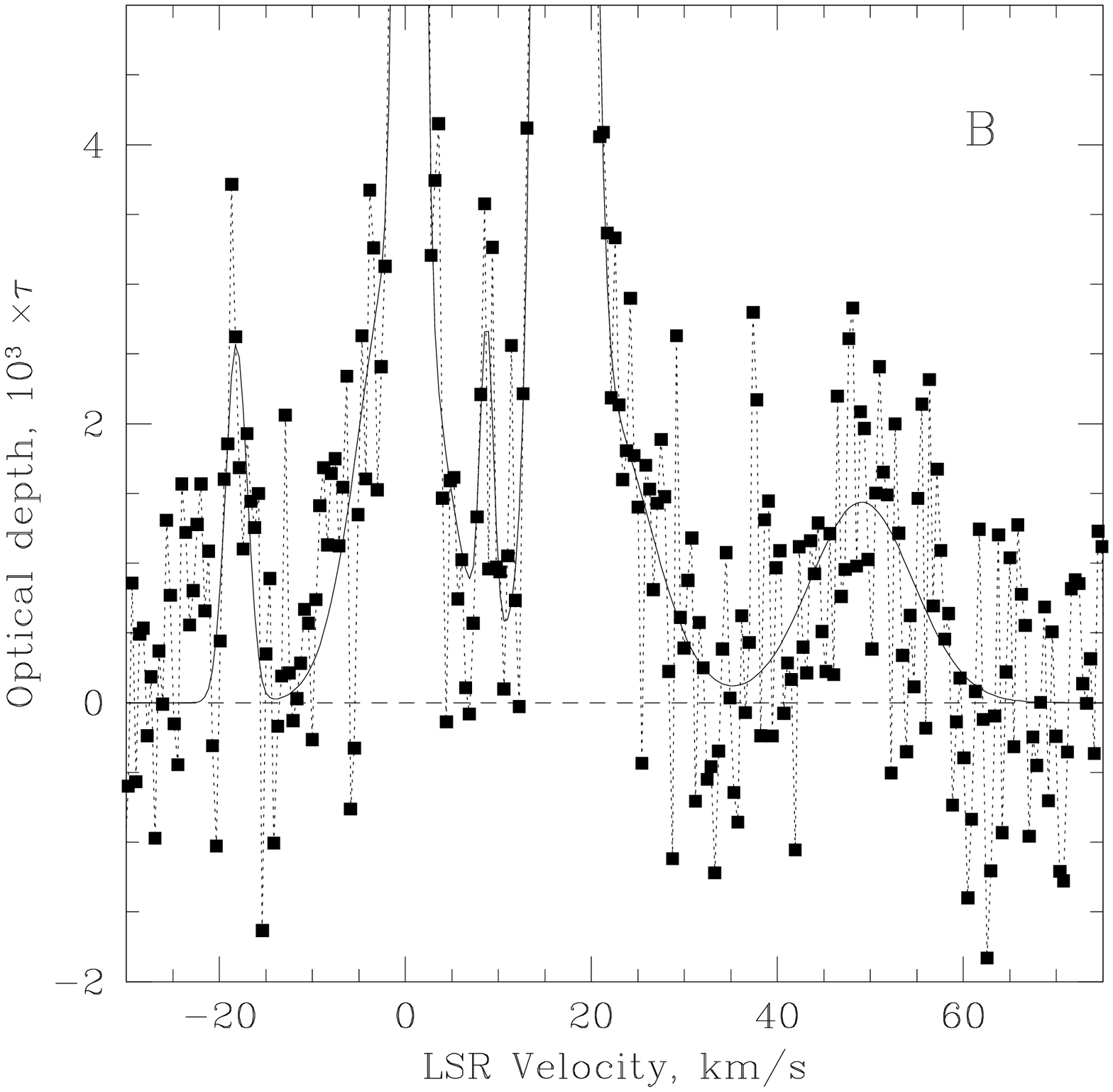,width=2.4in,height=2.4in}
\vskip -2.4 in
\hskip 4.5 in \epsfig{file=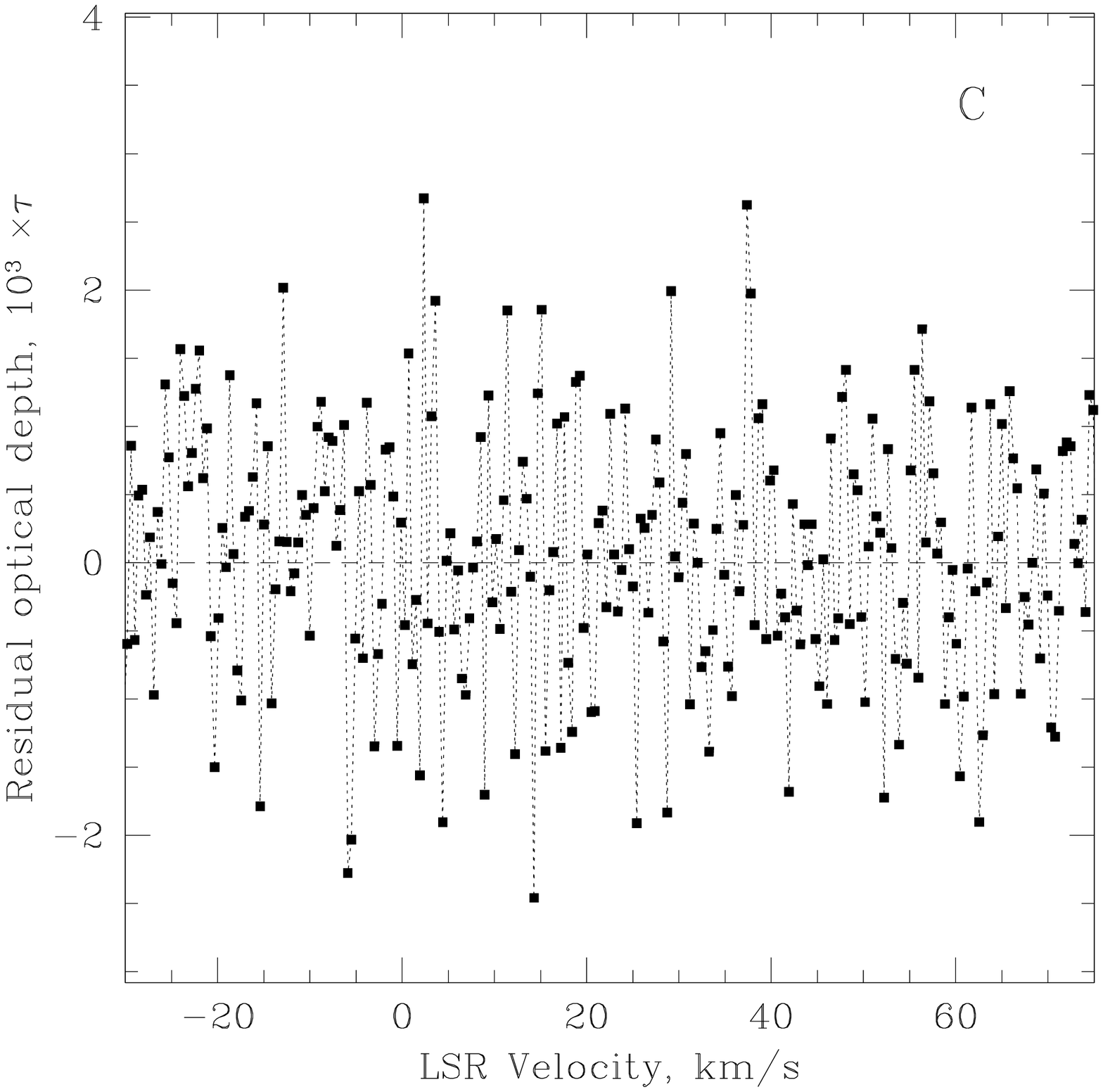,width=2.4in,height=2.4in}
\caption{ [A] : 0.4~\kms~resolution absorption spectrum toward \PKSB~(solid 
		points), along with the 8-Gaussian fit (solid line). 
          [B] : Zoomed-in version in which wide absorption can be seen at 
		the base of the narrow CNM features as well as at an LSR velocity 
		of $\sim +49$~\kms.
          [C] : The residual absorption spectrum after subtracting the fit; 
		the residuals are seen to lie within the noise.}
\label{fig:0407}
\vskip -0.05in
\end{figure*}

\PKSA~ and \PKSB~ were observed using the 6.0A array configuration of the ATCA 
in January 2001. This has a shortest baseline of 337~m, i.e.  $\sim 1.7$~k$\lambda$ 
at the 21~cm line frequency, ensuring that foreground HI 
emission is resolved out and does not affect the HI absorption spectrum. 
The system was configured to observe dual linear polarizations with a 4~MHz 
bandwidth covered by 2048 independent frequency channels; this provided a 
total velocity coverage of $844$~\kms~and a channel spacing of $0.4$~\kms. 
Observations of PKS~1934$-$638 were used to calibrate the absolute flux density 
scale. The passband was calibrated through frequency switching every fifteen 
minutes on the target sources (which are themselves phase calibrators for 
the ATCA): the off-line scans were offset in frequency from the on-line 
ones by 10~MHz in order to implement the frequency switching purely as 
frequency changes in the first local oscillator, while maintaining the 
conversion chain electronics unchanged. This observing strategy was developed to
obtain spectral baselines of high quality as the observations were aimed 
at detecting weak, broad spectral features. The total on-source, on-line time was 
five hours on each source.

The data were analyzed in {\sc MIRIAD}. After the initial editing, gain and 
bandpass calibration and continuum subtraction, the data were shifted to the 
LSR velocity frame and spectral cubes were made from the continuum-subtracted 
visibilities. No absorption was detected toward other sources in the two fields;
deconvolution was hence unnecessary and the final spectra toward the target sources 
were obtained by taking a cut through the cubes at the source positions. 
The flux densities of \PKSA~ and \PKSB~ were measured to be 14.38~Jy and 
16.16~Jy respectively, while the spectra each have an RMS noise of $\sim 0.001$ 
in units of optical depth (per $0.4$~\kms~ spectral channel). 
The analysis was repeated after rejecting the shortest baseline (337~m), to test 
whether the spectra were at all affected by emission structure in this 
baseline; the shortest baseline in the remaining data was 628~m (i.e. 
$\sim 3.2$~k$\lambda$).  The spectra were found to be unchanged, apart 
from a marginal increase in the noise.

\vskip -0.1in
\section{Results and Discussion}
\label{sec:results}

\subsection{Spectra}
\label{sec:spectra}

The final 0.4~\kms~resolution absorption spectrum toward \PKSA~is shown in 
Fig.~\ref{fig:1814}(A) (solid points); here, optical depth is plotted against 
LSR velocity. For clarity, only the central 100~\kms~of the spectrum are shown. 
Fig.~\ref{fig:1814}(B) shows a zoomed-in version of this spectrum; shallow wide 
absorption can be clearly seen over the velocity range $-10$ to $+10$~\kms. Since 
the deep narrow component is clearly asymmetric and there is additional broad 
absorption, we attempted to simultaneously fit three Gaussians to the absorption 
profile. This yielded an extremely stable, good fit, shown in Figs.~\ref{fig:1814}(A) 
and (B) as a solid line. Attempts were also made to fit only two Gaussians to the profile 
but these were found to leave large residuals. Fig.~\ref{fig:1814}(C) shows the 
spectrum after subtracting out the 3-component fit; the residuals are seen to 
lie within the noise. This residual spectrum was then smoothed to various 
velocity resolutions, up to a resolution of $35$~\kms, to search for additional 
broad absorption; no new components were detected. 

The absorption spectrum of \PKSB, shown in Fig.~\ref{fig:0407}(A), is far more 
complex than that of \PKSA. 
Fig.~\ref{fig:0407}(B) shows a zoomed-in version 
of the spectrum to emphasize the numerous absorption features spread over the 
range $-30$ to $+60$~\kms; again, the solid points show the measured optical 
depth while the solid line shows our best multi-Gaussian fit. Wide, weak 
absorption features can again be seen in this spectrum, both at the base of 
the strong CNM components (velocity range : $-15$ to $+30$~\kms) and, 
interestingly, at a location away from the central absorption complex, 
at an LSR velocity of $\sim +49$~\kms. Eight Gaussians were needed to 
obtain a good fit to the complex absorption profile. Attempts were again 
made to fit fewer Gaussians to the profile but these left large residuals in all 
cases. Fig.~\ref{fig:0407}(C) shows the residuals after subtracting out 
the eight-Gaussian fit from the spectrum. These are seen to lie within the noise;
no further components were detected on smoothing to coarser resolutions 
(up to $20$~\kms). While the large number of components needed to obtain 
a good fit does raise questions about the uniqueness of the decomposition, 
the fit to the $+49$~\kms~component is likely to be a good one as this 
is shifted in velocity relative to the main absorption components and the 
fit is thus only marginally affected by them. 

\subsection{HI kinetic temperatures}
\label{sec:temp}

\setcounter{table}{0}
\begin{table*}
\label{table:gauss} 
\begin{centering}
\caption{Parameters of the simultaneous multiple Gaussian fits to the absorption spectra.}
\begin{tabular}{|c|c|c|c|c|c|c|}
\hline
Source & Component  & Optical depth   & LSR velocity & FWHM   & $T_{\rm k}$ & N$_{\rm HI}$ \\
       &            &$\tau_{\rm max}$ &  (km/s)      & (km/s) &   K   & $\times 10^{20}$~\cm \cr
\hline
      & 1 & $0.306 \pm 0.003$ & $-0.903 \pm 0.002$ & $1.43 \pm 0.01$ & $44.6 \pm 0.7$ & $0.38 \pm 0.01$ \cr
\PKSA & 2 & $0.110 \pm 0.003$ & $-1.05 \pm 0.01$   & $3.37 \pm 0.06$ & $248 \pm 10$   & $1.77 \pm 0.15$ \cr
      & 3 &$0.0099 \pm 0.0008$& $-2.34 \pm 0.20$   & $12.0 \pm 0.5$  & $3127 \pm 300$ & $7.2 \pm 1.7$   \cr
\hline
      & 1 & $0.111 \pm 0.002$ & $16.394 \pm 0.008$ & $1.73 \pm 0.03$ & $65.7 \pm 2.1$ & $0.24 \pm 0.02 $ \cr
      & 2 & $0.042 \pm 0.002$ & $17.03 \pm 0.05$   & $3.94 \pm 0.12$ & $339 \pm 21$   & $1.1 \pm 0.2 $ \cr
      & 3 &$0.0255 \pm 0.0007$& $0.51 \pm 0.03 $   & $2.11 \pm 0.08$ & $97.2 \pm 7.2$ & $0.10 \pm 0.02$ \cr
\PKSB & 4 &$0.0014 \pm 0.0002$ & $49.05 \pm 1.08 $ & $13.0 \pm 2.6 $ & $3694 \pm 1595$& $1.3 \pm 1.2 $ \cr
      & 5 &$0.0026 \pm 0.0005$ &$-18.18 \pm 0.28 $ & $2.8 \pm 0.7 $  & $165 \pm 88$   & $0.02 \pm 0.03 $ \cr
      & 6 &$0.0034 \pm 0.0005$ & $-0.51 \pm 0.46 $ & $10.0 \pm 1.3 $ & $2165 \pm 608$ & $1.4 \pm 0.9 $ \cr
      & 7 &$0.0022 \pm 0.0005$ & $20.7 \pm 1.5 $   & $12.7 \pm 2.2 $ & $3500 \pm 1354$& $1.9 \pm 1.9 $ \cr
      & 8 &$0.0023 \pm 0.0007$ & $8.75\pm 0.25$    & $1.6 \pm 0.6$   & $59.3 \pm 54.7$& $0.04 \pm 0.13$ \cr
\hline
\end{tabular}
\end{centering}
\end{table*}

Table~\ref{table:gauss} lists the parameters of the multiple Gaussian fits to 
the two optical depth spectra. Here, Col.~3 gives the peak optical depth in 
each component, Col.~4, the velocity location of this peak and Col.~5, the FWHM of 
the component. The last two columns are the kinetic temperature and HI column 
density of the component, obtained from the expressions $T_{\rm k} = 
21.855 \times {\Delta V}^2$ and $\NHI = 1.823 \times 10^{18} \times T_{\rm k} \times 
[ 1.06 \times \tau_{\rm max} \times \Delta V ]$, where $\Delta V$ is the 
FWHM in \kms~and we have used $\int \tau dV =  1.06 \times \tau_{\rm max} 
\times \Delta V $ (valid for a Gaussian) and assumed $T_{\rm s} = T_{\rm k}$. 
It should be emphasized that the above expression for $\NHI$ is valid for 
the CNM but may not be valid for the WNM, where $T_{\rm s}$ may be lower 
than $T_{\rm k}$~\citep{liszt01}; the quoted column densities for the
WNM components should hence be viewed as upper limits. The above equations 
also assume that the observed velocity widths arise from Doppler broadening 
due to thermal motions; in the case of turbulent motions or blending of 
components, the values in Col.~6 and Col.~7 are upper limits on the 
true kinetic temperature and the HI column density.

In the case of \PKSA, two of the three components have temperatures close 
to the known CNM range, with $T_{\rm k_1} = 44.6 \pm 0.7$~K and 
$T_{\rm k_2} = 248 \pm 10$~K.  The third component has a velocity width 
(FWHM) of 12.0~\kms, implying a kinetic temperature $T_{\rm k_3} = 3127 
\pm 300$~K.  This lies significantly above the range of temperatures theoretically 
allowed for the CNM; however, it is below the canonical WNM range and in the 
thermally unstable range of temperatures $500 - 5000$~K \citep{wolfire95}. 
Thus, either HI indeed exists at thermally unstable temperatures in the Galaxy or 
the third component is due to non-thermally broadened CNM absorption. In the 
latter case, absorption by any WNM along this line of sight would be even weaker 
(and possibly broader) than this third component. However, after subtracting 
out the above three components and smoothing the spectrum to coarser resolutions, 
we find no evidence for any additional absorption. This non-detection places a 
$3\sigma$  upper limit of $1.6 \times 10^{20}$~\cm~on the column density of HI 
gas at a {\it spin temperature} of $8000$~K (i.e.  with $T_{\rm k} \ge 8000$~K; 
\citet{liszt01}). 
Note that this constraint on the column density is even stronger for a lower 
WNM spin temperature as the limit is directly proportional to $T_{\rm s}$. If 
the third component is indeed non-thermally broadened CNM at a kinetic temperature 
$T_{\rm k_3}$, its column density is N$_3 = 0.23 \times 10^{20} \times T_{\rm k_3}$~\cm~ 
(as $T_{\rm s} = T_{\rm k}$~ for the CNM).  Since $T_{\rm k_3} \gtrsim 40$~K 
\citep{wolfire95}, the lower limit to the total CNM column density along this line 
of sight (from all three components) is $2.4 \times 10^{20}$~\cm. Combining this 
with the above upper limit on the WNM column density yields a $3\sigma$ upper 
limit of $\sim 40$\% on the fraction of HI along this line of sight that is 
in the WNM. This contrasts with the picture that the CNM and WNM are equitably 
distributed (e.g. \citealt{kulkarni88}; note that Heiles \& Troland (2003b) 
find that as much as $\sim 60$\% of all HI is in the WNM phase). Further, 
\PKSA~is at a relatively high Galactic latitude; one would hence expect an 
even higher WNM fraction here than in lines of sight in the plane (e.g. 
\citealt{heiles03b}), due to the lower pressure away from the plane 
\citep{wolfire95}. On the other hand, if the kinetic temperature is indeed 
$\sim 3127$~K, the HI fraction in the WNM is $\lesssim 75$\%, as might be expected 
for a line of sight away from the plane. These arguments favour an interpretation 
where the third component arises in the WNM, with $T_{\rm k_3} = 3127$~K, in the 
thermally unstable range. 

Next, for \PKSB~, five of the Gaussian components in Table~\ref{table:gauss} 
are CNM, with temperatures in the range $T_{\rm k} \sim 60 - 340$~K. 
The remaining three components have kinetic temperatures $T_{\rm k} \sim 2100 - 3700$~K
and in the thermally unstable range.   Two of these three wide 
components (\#6 and \#7, at LSR velocities $-0.5$ and $20.7$~\kms~respectively) 
lie within the velocity range of CNM absorption; consequently, the fits to 
these components might be confused by the CNM features. 
Component \#4 (at $V_{\rm LSR} \sim 49$~\kms) is, however, some distance away 
from the central absorption complex; the fit to this component is thus 
likely to be unique. The temperature of this component is estimated to be 
$T_{\rm k} = 3694 \pm 1595$~K, from its velocity width. The errors on the fit are 
somewhat larger than that toward \PKSA, due to the complexity of the 
absorption spectrum and, consequently, the number of components needed to obtain 
a good fit. Unlike the case of \PKSA, it is difficult to constrain the 
WNM column density along this line of sight --- by adopting the hypothesis 
that all the wide absorption components are non-thermally broadened CNM --- 
due to the possibility that a broad warm component might be lost in the 
welter of features in the central absorption complex.

It would be interesting to estimate spin temperatures by the ``classical'' 
method and to compare them to the kinetic temperatures of Table~\ref{table:gauss}.
One could also use the absorption fits to ``predict'' the emission profile and 
compare this to the observed emission. While both of these would serve as
cross-checks to the derived parameters, one should note that modelling
the emission profile requires additional assumptions about the distribution 
of the CNM and WNM along the line of sight. This is especially critical for 
velocity regions containing multiple components, such as the central absorption 
complex towards \PKSB. Next, while the present observations also allowed us 
to measure the HI emission profiles along the two lines of sight, the large 
primary beam of the 22m ATCA dishes imply that  these spectra are very likely
to be affected by the issues discussed in Section~\ref{sec:intro}. With
this caveat in mind, we will use the present spectra for a brief comparison
between $T_{\rm s}$ and $T_{\rm k}$ and the predicted and observed emission profiles.
In the case of the +49~\kms~component towards \PKSB, reasonable agreement 
can be obtained between the predicted and observed emission if we assume 
$T_{\rm s} \sim 1400$~K, i.e.  a factor of $\sim 2.5$~less than the kinetic 
temperature. Unfortunately, the large errors on $T_{\rm k}$ imply that this
estimate of $T_{\rm s}$ is, in fact, within $1.5\sigma$ of the kinetic
temperature; the above comparisons are thus not very meaningful along this
line of sight. We note, further, that  our ATCA emission spectrum toward 
\PKSB~shows about twice the brightness temperature seen in the Parkes spectrum 
of \citet{rad72}; this suggests that the HI has structure within the primary 
beam of the AT dishes, making a comparison between the observed and derived 
emission profiles unreliable. Observations are presently being carried 
out to obtain HI emission mosaics in the vicinity of both sources, with 
both high spatial and spectral resolution; these will be used to redo
the above comparisons in detail. We hence defer a full comparison along this line 
of sight until these mosaic images are available.
On the other hand, in the case of the relatively simple line
of sight toward \PKSA, $T_{\rm s}$ is found to be in reasonable agreement
with $T_{\rm k}$, apart from the velocity range $-6$ to $+3$~km/s where CNM absorption
contributes significantly to the absorption profile (and thus lowers the spin 
temperature). This supports the 
argument that the third absorption component indeed arises in the WNM. It is 
also interesting to note that the estimated $T_{\rm s}$ values are, in general, again
somewhat {\it lower} than the kinetic temperature $T_{\rm k} = 3127 \pm 300$~K.
On the other hand, the predicted emission profile (assuming no absorption of 
background emission by cold foreground HI) has a higher peak brightness temperature 
than that observed in our ATCA spectrum by about a factor of three; the observed 
emission spectrum also has wider wings than the model emission profile. It is, at 
present, unclear if this is due to a significantly warmer undetected WNM phase 
(with $T_{\rm k} \gtrsim 10^4$~K) or because of the low angular resolution of the 
emission profile or emission-related issues. We note, finally, that it is possible to 
obtain a far better agreement between the predicted and observed emission profiles 
by leaving the WNM spin temperature and the amount of absorption of background 
emission by foreground CNM as free parameters. However, we again defer a full 
analysis till the mosaic emission profiles are available.

\subsection{Discussion}
\label{sec:discuss}

In recent times, deep searches have been carried out for the WNM using both 
interferometers \citep{carilli98,dwaraka02} and single dishes 
\citep{heiles03a,heiles03b}, again via a comparison between 
absorption and emission spectra. Carilli et al. (1998) used the Westerbork 
Synthesis Radio Telescope (WSRT) to detect weak ($\tau_{21} \la 10^{-3}$)
broad absorption toward Cygnus~A, blended with numerous, much deeper 
CNM features. They identified the broad component with the WNM,
obtaining $T_{\rm s} \sim 6000 \pm 1700$~K and $T_{\rm s} \sim 4800 \pm 
1600$~K in two velocity ranges (in broad agreement with the earlier 
single-dish results of \citet{mebold75} and \citet{kalberla80}).
Similarly, \citet{dwaraka02} detected wide absorption toward 
3C147 with the WSRT and estimated $T_{\rm s} \sim 3600 \pm 360$~K. 
However, both these studies could be affected by the problems of comparing 
on-source absorption spectra with off-source emission. The observations 
also had relatively poor velocity resolution ($\sim 2.1$~\kms), allowing 
the possibility that the observed broad absorption is a blend of 
narrow CNM lines.  It should be emphasized that these studies yielded 
estimates of the WNM {\it spin} temperature, which, as discussed earlier, may 
be lower than the kinetic temperature.

On the other hand, Heiles \& Troland (2003a,b) used the Arecibo Telescope 
to carry out high velocity resolution ($\sim 0.4$~\kms) 21~cm 
absorption/emission studies toward a number of compact radio sources; 
the high spectral resolution allowed them to fit Gaussians to the narrow 
CNM absorption features and to then model the emission spectra as a 
sum of the CNM Gaussians and additional Gaussians from the WNM. A 
least-squares fit to the emission spectra was then used to estimate 
the {\it kinetic} temperature of WNM components. A substantial fraction 
($\sim 48$\%) of the WNM was found to be in the thermally unstable phase, with 
kinetic temperatures in the range $\sim 500 - 5000$~K.
While these results are exceedingly interesting, the observations were single-dish ones 
and hence subject to the problems discussed earlier. Attempts were made to correct 
for some of these issues by (1)~using a grid of off-source pointings to better 
constrain the on-source emission spectrum by estimating the spatial derivatives of 
the brightness temperature in different directions and (2)~including the 
effects of self-absorption in the least-squares fit. As the authors mention,
the latter was indeterminate in most cases and it was hence only possible 
to distinguish between extreme situations. The effect of this uncertainty 
on their results is not well understood.

The present approach essentially combines the good features of both the 
above methods, using high spectral resolution and interferometric baselines; 
the crucial difference is that we work entirely with the absorption 
spectra and are thus not affected by emission-related issues. The critical 
assumption involved in our analysis is the decomposition of the absorption 
profiles into thermally broadened Gaussians. If this assumption breaks 
down (e.g. due to blending of narrower components), 
our estimates provide upper limits on the kinetic temperature for the different 
components.

Wide absorption was detected along both lines of sight discussed here, with 
four components showing kinetic temperatures in the thermally unstable range $2000 
< T_{\rm k} < 5000$~K.  The results appear quite robust for the line of sight 
toward \PKSA, due to the relative simplicity of the absorption profile. 
Similarly, while the fits to the two central wide components (\#6 and \#7 
in Table~\ref{table:gauss}) toward \PKSB~ may not be unique, the component 
at $+49$~\kms~LSR velocity appears to be well fit by a single Gaussian. 
There is also no evidence suggesting that the four wide components are 
non-thermally broadened CNM; moreover, in the case of \PKSA, the deduced 
CNM fraction and spin temperatures support the case that the wide absorption 
arises from the WNM. It thus appears that there do exist WNM components with 
kinetic temperatures in the thermally unstable range, in agreement with the 
earlier results of \citet{heiles03a,heiles03b}. This indicates that thermal 
equilibrium has not been reached throughout the WNM, possibly due to the 
low number densities here and hence the long time-scales needed to reach 
equilibrium (e.g. \citealt{wolfire03}). New and upgraded radio interferometers 
such as the ATCA, the WSRT and the Giant Metrewave Radio Telescope will allow 
this hypothesis to be tested on a statistically significant number of lines 
of sight in the Galaxy, thus enabling us to arrive at a better understanding 
of this important phase of the interstellar medium. 
\vskip 0.05in
\noindent {\bf Acknowledgments} The Australia Telescope is funded by the Commonwealth of 
Australia for operation as a National Facility managed by CSIRO.

\label{lastpage}

\end{document}